\documentclass{ptapap}
\usepackage{color}

\usepackage{soul}

\author{Dawid Moździerski}[IAUWr]
\author{Andrzej Pigulski}[IAUWr]
\author{Zbigniew Kołaczkowski}[IAUWr,CAMK]
\author{Gabriela Michalska}[IAUWr]
\author{Grzegorz Kopacki}[IAUWr]
\author{Artur Narwid}[IAUWr]
\author{Marek Stęślicki}[CBK]
\author{Ewa Zahajkiewicz}[IAUWr]
\author{Jianning Fu}[BNU]
\author{Xiaojun Jiang}[NAO]
\author{Chao Zhang}[BNU]
\author{Jason Jackiewicz}[NMSU]
\author{John Telting}[NOT]
\author{Thierry Morel}[IAG]
\author{Piotr Śródka}[IAUWr]
\author{Przemysław Bruś}[IAUWr]
\author{Fabien Carrier}[KUL]
\affil[IAUWr]{Astronomical Institute, University of Wroc{\l}aw, Wroc{\l}aw, Poland}
\affil[CAMK]{Nicolaus Copernicus Astronomical Center, Warszawa, Poland}
\affil[CBK]{Space Research Centre, Polish Academy of Sciences, Wrocław, Poland}
\affil[BNU]{Department of Astronomy, Beijing Normal University, Beijing, China}
\affil[NAO]{National Astronomical Observatories, Chinese Academy of Sciences, Beijing, China}
\affil[NMSU]{New Mexico State University, Las Cruces, NM, USA}
\affil[KUL]{Katholieke Universiteit Leuven, Instituut voor Sterrenkunde, Leuven, Belgium}
\affil[NOT]{Nordic Optical Telescope, La Palma, Spain}
\affil[IAG]{Institut d'Astrophysique et de G\'{e}ophysique, Universit\'{e} de Li\`{e}ge, Li\`{e}ge, Belgium}

\title{Variability  in NGC\,6910, the open cluster rich in $\beta$\,Cephei-type stars}

\begin{document}

\maketitle

\begin{abstract}

NGC\,6910 is the northern hemisphere open cluster known to be rich in $\beta$\,Cephei-type stars.
Using four-season photometry obtained in Bia\l k\'ow (Poland) and Xinglong (China) observatories,
we performed variability survey of NGC\,6910. As the result, we found over 100 variable stars in the
field of the cluster, including many stars showing variability due to pulsations and binarity. Thanks to the spectroscopic observations, we also detected changes in the profiles of spectral lines of $\beta$\,Cep
stars, caused by pulsations.

\end{abstract}

\section{Introduction}

NGC\,6910 is the young open cluster containing many
$\beta$\,Cep-type variables \citep{Zibi2004}. Preliminary results of the variability search
based on photometric data obtained during the international observational campaign
allowed to detect eight $\beta$\,Cep-type members \citep{Pigu2008, Saes2010}. Interestingly,
it turned out that the frequency spectra of these $\beta$\,Cep stars, arranged according to the 
decreasing brightness (i.e.~mass) showed a progression of frequencies of the excited modes \citep{Pigu2008}.
This result raised hope for a successful ensemble asteroseismology in this cluster \citep{Saes2010}. In
the present paper, we show preliminary results of the full
variability survey of NGC\,6910 based on the part of the data obtained during the international
observational campaign in the years 2005 -- 2007 and 2013. The full results will be published elsewhere.

\section{Observations and Results}

\begin{figure}
\begin{center}
\includegraphics[width=8cm]{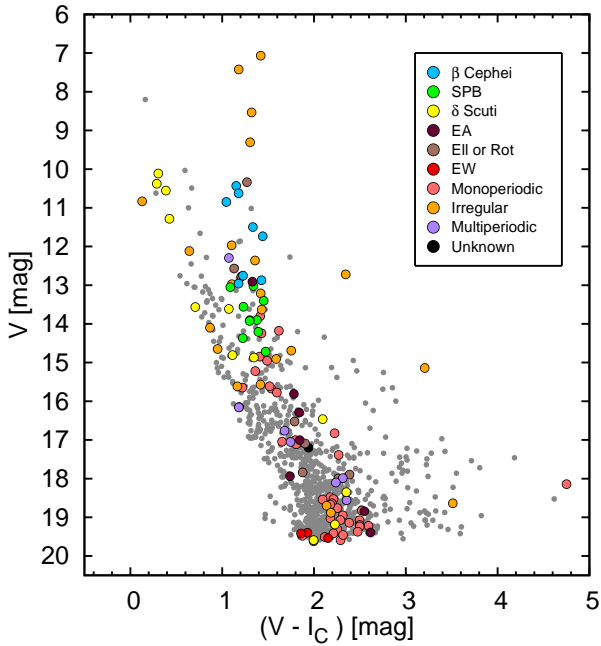}
\caption{Colour--magnitude diagram for NGC\,6910 open cluster. Variable stars are marked by coloured circles.}
\label{Var_CMD}
\end{center}
\end{figure}

We used only Bia\l k\'ow (Poland) and Xinglong (China) photometric data, the two most numerous data samples.
Bia\l k\'ow observations were carried out with a 60--cm reflecting
telescope and the attached CCD camera covering $13^\prime \times12^\prime$ field of view
(FoV). Xinglong data
were carried out with 50 and 100--cm reflecting telescopes, and CCD cameras covering,
$10^\prime\!\!.5\times10^\prime\!\!.4$ and $11^\prime\!\!.2\times11^\prime\!\!.2$ FoV, respectively. In total, about 23\,500 CCD frames were acquired
through the $B$, $V$, $R$, $I_{\rm C}$ bands and narrow--band H$\alpha$ filters during 157 nights in these two sites.
We also carried out
spectroscopic echelle observations with the 2.56--m Nordic Optical Telescope, 1.93--m telescope
of Haute--Provence Observatory and Apache Point Observatory 3.5--m Astrophysical Research Consortium telescope in 2007
and 2013.

Photometric observations allowed us to detect 125 variable stars in the field of NGC\,6910
open cluster, of which 117 are new (Fig.~\ref{Var_CMD}, Tab.~\ref{tab1}). Thanks to the large
difference in longitude between both observatories, we were able to
reduce significantly the daily aliases in the frequency spectra. One of the most
interesting results 
is the mentioned earlier progression of the frequencies for $\beta$\,Cep stars, up to almost $13\,{\rm d}^{-1}$
in NGC6910-38. We also detected changes of amplitudes of some modes in two $\beta$ Cep stars,
NGC6910-16 and NGC6910-27 between 2005 -- 2007 and 2013.

Thanks to the spectroscopic
observations we have found well pronounced variability of spectral line profiles caused by pulsations for
two $\beta$\,Cep stars: NGC6910-14 and NGC6910-18. We will use this information to identify the
degrees $l$ and azimuthal orders $m$ of the strongest modes using methods of \cite{Zima2006}, \cite{Dasz2005} and \cite{Dasz2009}.

\begin{table}
\begin{center}
\caption{Variable stars in the observed FoV and probable NGC\,6910 membership.}
\label{tab1}
\vspace{5mm}
\small{
\begin{tabular}{c|c|c}
\textbf{Variability} & \textbf{Likely} & \textbf{Non--}  \\
\textbf{type}        & \textbf{members}& \textbf{members}\\
\hline
$\beta$\,Cep   &  8 &  0 \\
SPB            & 10 &  0 \\
$\delta$ Sct   &  0 & 12 \\
EA             &  3 &  4 \\
Ell or Rot     &  2 &  6 \\
EW             &  0 &  3 \\
Monoperiodic   & 34 & 12 \\
Irregular      & 16 &  6 \\
Multiperiodic  &  0 &  7 \\
Unknown        &  0 &  2 \\
\hline
\textbf{Total} & \textbf{73} & \textbf{52} \\ 
\end{tabular}
}
\end{center}
\vspace{1mm}
\end{table}

\acknowledgements{This work was supported by the Polish National Science Center grants No. 2016/21/B/ST9/ 01126, No. 2012/05/N/ST9/03898
and has received funding from the European Community’s Seventh Framework Programme (FP7/2007-2013) under
grant agreement no. 269194. This research has made use of the WEBDA database, operated at the Department
of Theoretical Physics and Astrophysics of the Masaryk University.}

\bibliographystyle{ptapap}
\bibliography{NGC6910vars}

\end{document}